\documentclass[slac_one]{revtex4}
\usepackage{graphicx}
\usepackage{fancyhdr}
\newcommand{\ttb}       {t$\bar{\textrm{t}}$}

\def\gevcc{GeV/c$^2$}
\def\gevc{GeV/c}
\def\tevcc{TeV/c$^2$}
\def\invfb{fb$^{-1}$}
\def\invpb{pb$^{-1}$}
\def\lowlumi{2$\times$10$^{33}$~cm$^{-2}$s$^{-1}$}

\def\dphi{$\Delta\eta$}
\def\pt{p$_{\rm{T}}$}
\def\et{E$_{\rm{T}}$}

\def\gamgam{$\gamma\gamma$}
\def\tautau{$\tau\tau$}
\def\rarr{$\rightarrow$}

\begin{document}

\title{{\small{Hadron Collider Physics Symposium (HCP2008),
Galena, Illinois, USA}}\\ 
\vspace{12pt}
Standard Model Higgs Searches at the LHC} 

\author{Maiko Takahashi (for the ATLAS and CMS collaborations)}
\affiliation{Imperial College London / University of Manchester, United Kingdom}

\begin{abstract}
An overview of the searches for the Standard Model Higgs boson at the LHC 
is presented. The main Higgs production and decay modes that have been studied 
are introduced, and the analysis techniques and the recent developments done 
by the ATLAS and CMS experiments are described. Some preliminary results from
current studies are included. The discovery potential within the first few 
years of physics running is evaluated.
\end{abstract}

\maketitle

\thispagestyle{fancy}

\section{Introduction} \label{sec:intro} 
The Large Hadron Collider (LHC) will enable production of the Standard
Model Higgs boson in the entire range of its allowed mass; from the
lower experimental exclusion limit of 114.4\gevcc~\cite{bib:LEP} to
the theoretical limit of $\sim$1\tevcc. An early discovery of the
Higgs boson is the primary objective of the two general purpose
detector experiment, ATLAS and CMS. The search requires a good
understanding of both the data and the physics of the background
processes, and extensive studies based on MC simulation are being
carried out by both experiments to prepare for the analysis of the LHC
data. Over the past years, the analyses have constantly been
improved. New event generators and more precise description of the
detector in the simulation are employed, and more sophisticated
analysis techniques and tools are being developed.

In this paper, an overview of the activities within the ATLAS and CMS
collaborations towards the discovery of the Higgs boson at the LHC is
presented. This includes basic ideas and the particular challenges, as 
well as some of the new developments and recently updated results for 
the main channels studied. Many of
the analyses are targeted towards the first year to 3 years of physics
runs where a low luminosity (L = \lowlumi) environment is
assumed. Both experiments are currently undertaking a major update of
the analysis results, many of which are unavailable at present,
however, new results are expected to be made public in the near future
before the arrival of the LHC physics data. The prospects for the
Higgs searches in the early years of LHC operation including all the
channels introduced here are summarised at the end.
 
\section{Higgs at the LHC}
\label{sec:Higgs} 
\begin{figure}[tbh!]
\begin{center}
\begin{tabular}{cc}
\includegraphics[angle=-90, width=.44\textwidth]{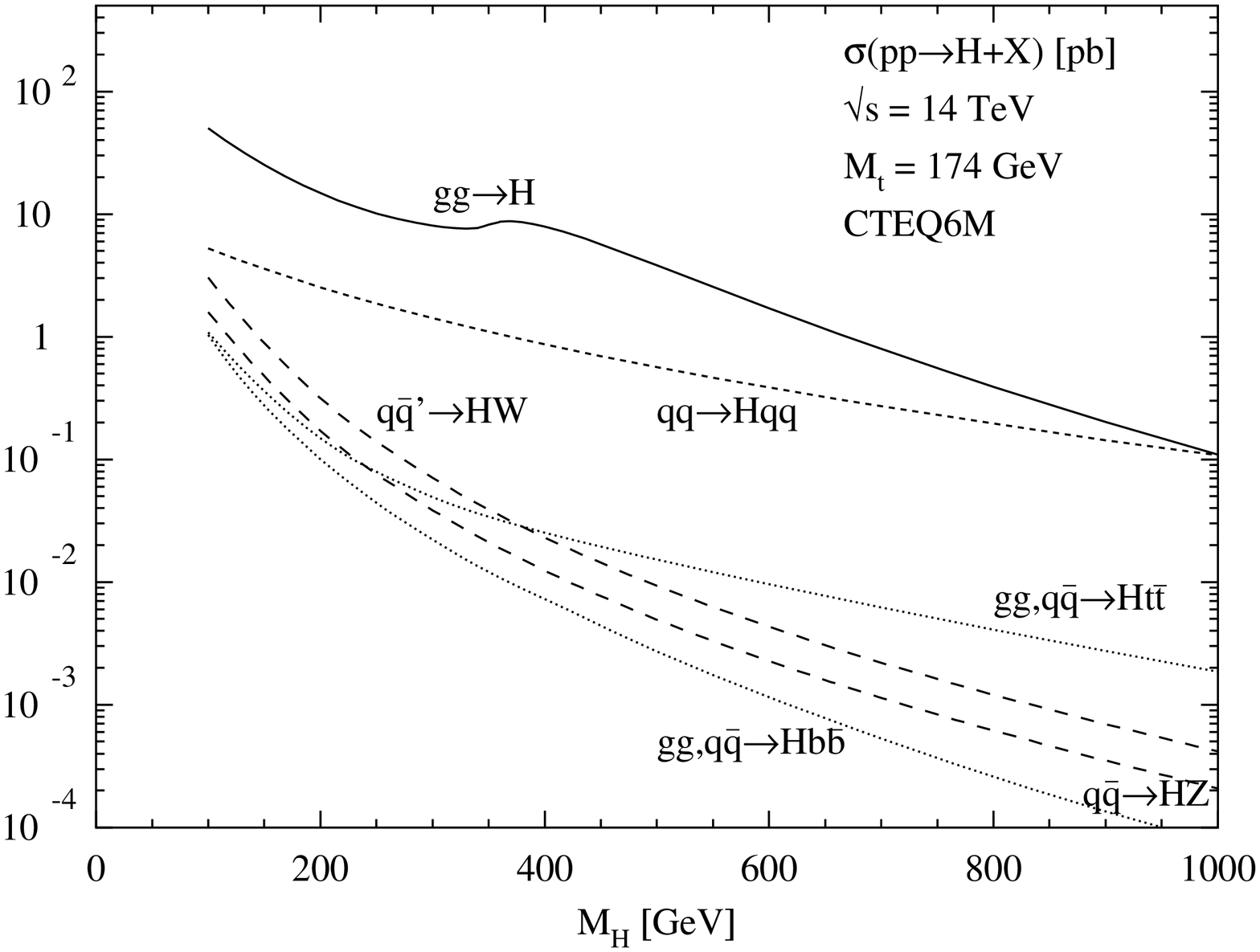} 
& \includegraphics[angle=-90, width=.44\textwidth]{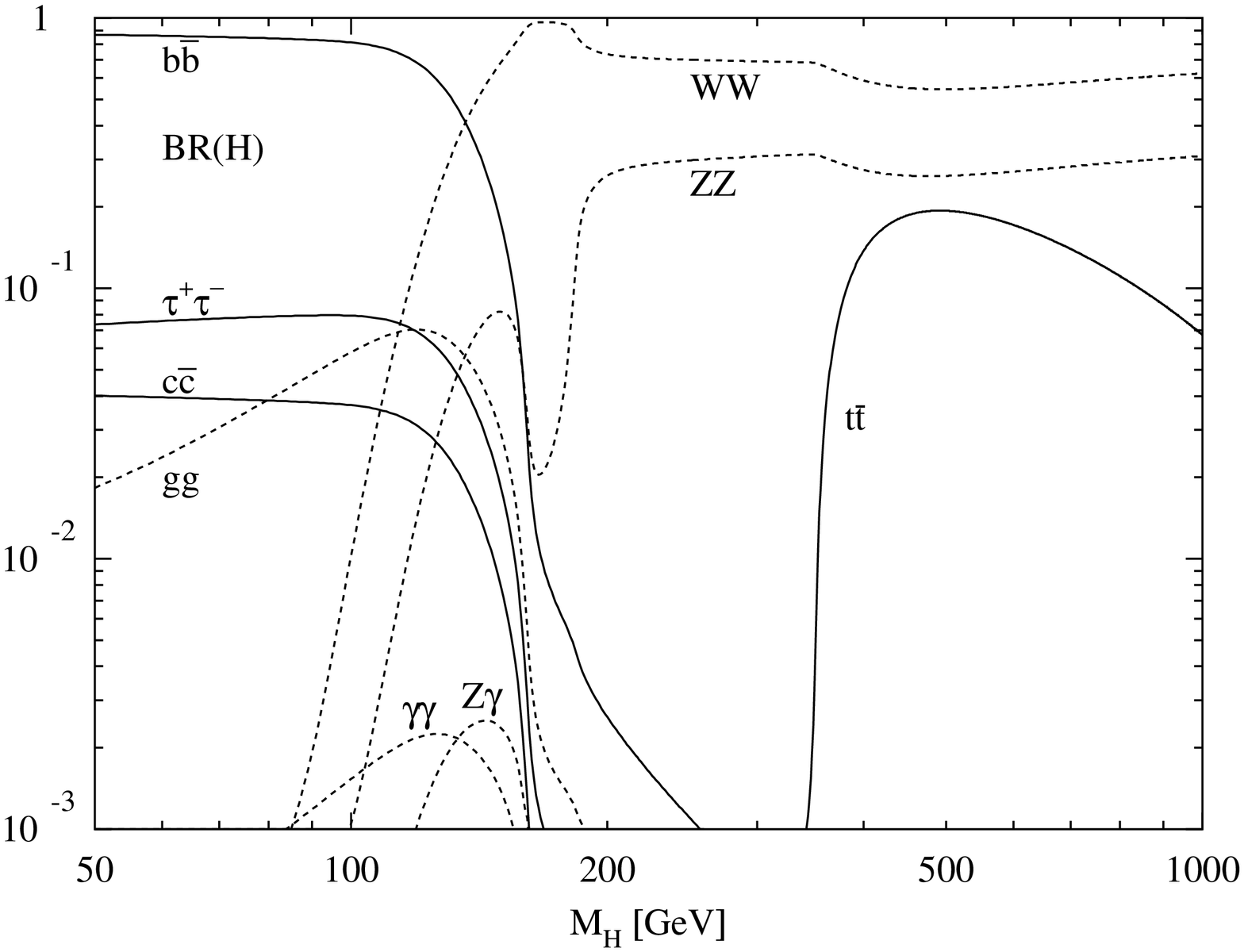} \\
&\\
(a) & (b) \\
\end{tabular}
\end{center}
   \caption
       { (a) The Standard Model Higgs boson production cross section
	 expected at the LHC and (b) the Higgs decay branching ratio
	 as a function of its mass.  }
\label{fig:Higgs}
\end{figure}

The most dominant mechanism for the production of the Higgs boson at
the LHC is the gluon fusion, followed by the Vector Boson Fusion (VBF)
which is approximately an order of magnitude smaller in cross section
(Fig.~\ref{fig:Higgs} (a)). While inclusive searches have the
advantage of having a large cross section, the VBF or other production
mode specific studies are also carried out by the ATLAS and CMS
experiments to improve the sensitivity of the search by exploiting the
additional signatures in the final state. If the Higgs boson mass is
below $\sim$130~\gevcc, the main decay modes are bb and \tautau~
(Fig.~\ref{fig:Higgs} (b)). These are also the challenging channels
due to the presence of hadronic decay and the relatively small
\pt~involved in the final state particles. The H\rarr\gamgam~decay
mode, on the other hand, provides a clean signature of high
\pt~photons which can be well reconstructed, hence plays an important
role in the Higgs searches at the LHC. For the higher Higgs mass
region, the decays to the vector bosons, W and Z, are the dominant
channels. Both ATLAS and CMS detectors were designed to deliver the
optimal performance for the observation of these benchmark channels.

\section{Low Mass Higgs Searches}
\label{sec:lowmass}

\subsection{H\rarr\gamgam}
\label{sec:Hgamgam}

\begin{figure}[tb!]
\begin{center}
\begin{tabular}{cc}
\includegraphics[width=.5\textwidth]{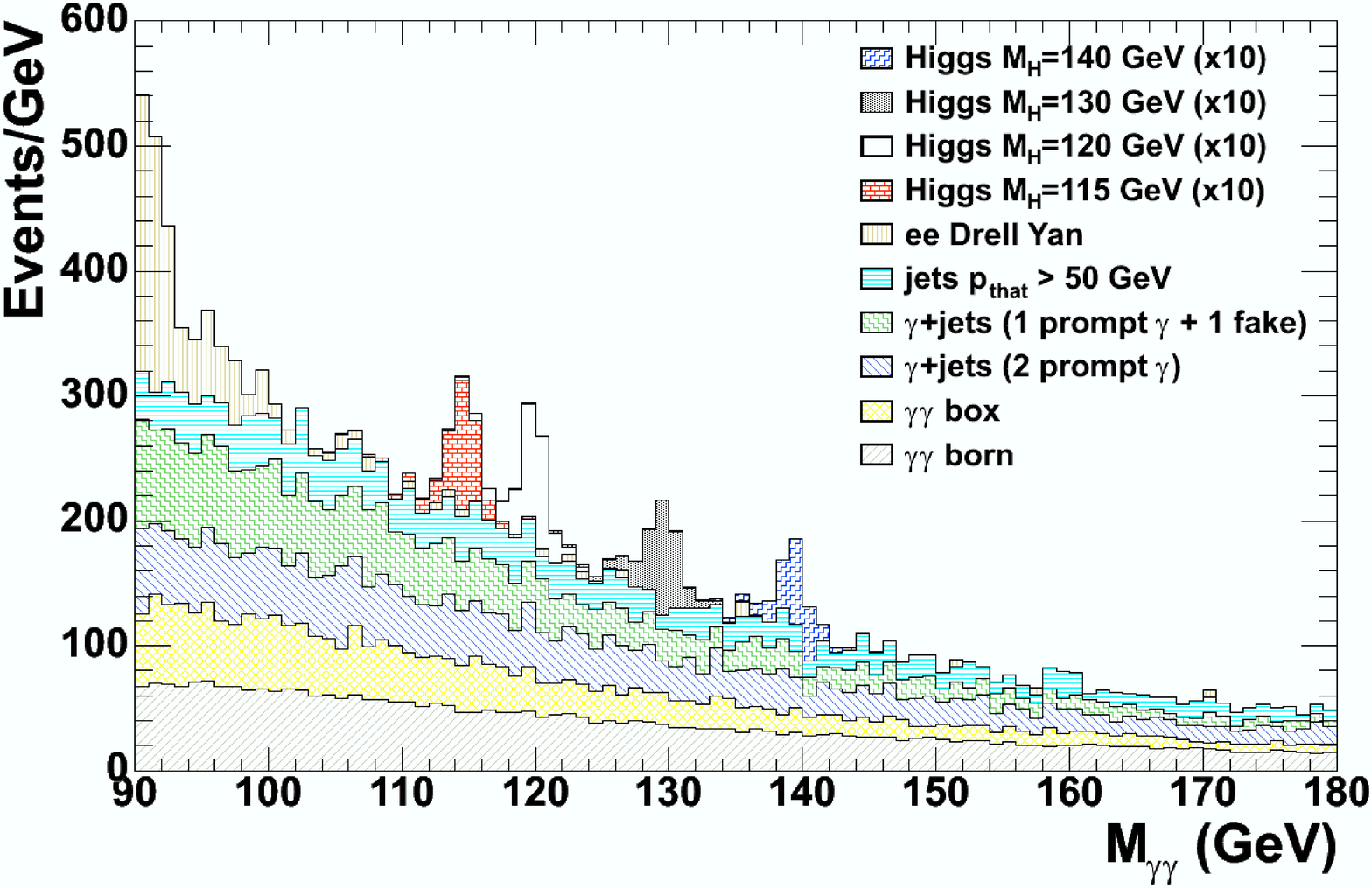} 
& \includegraphics[width=.5\textwidth]{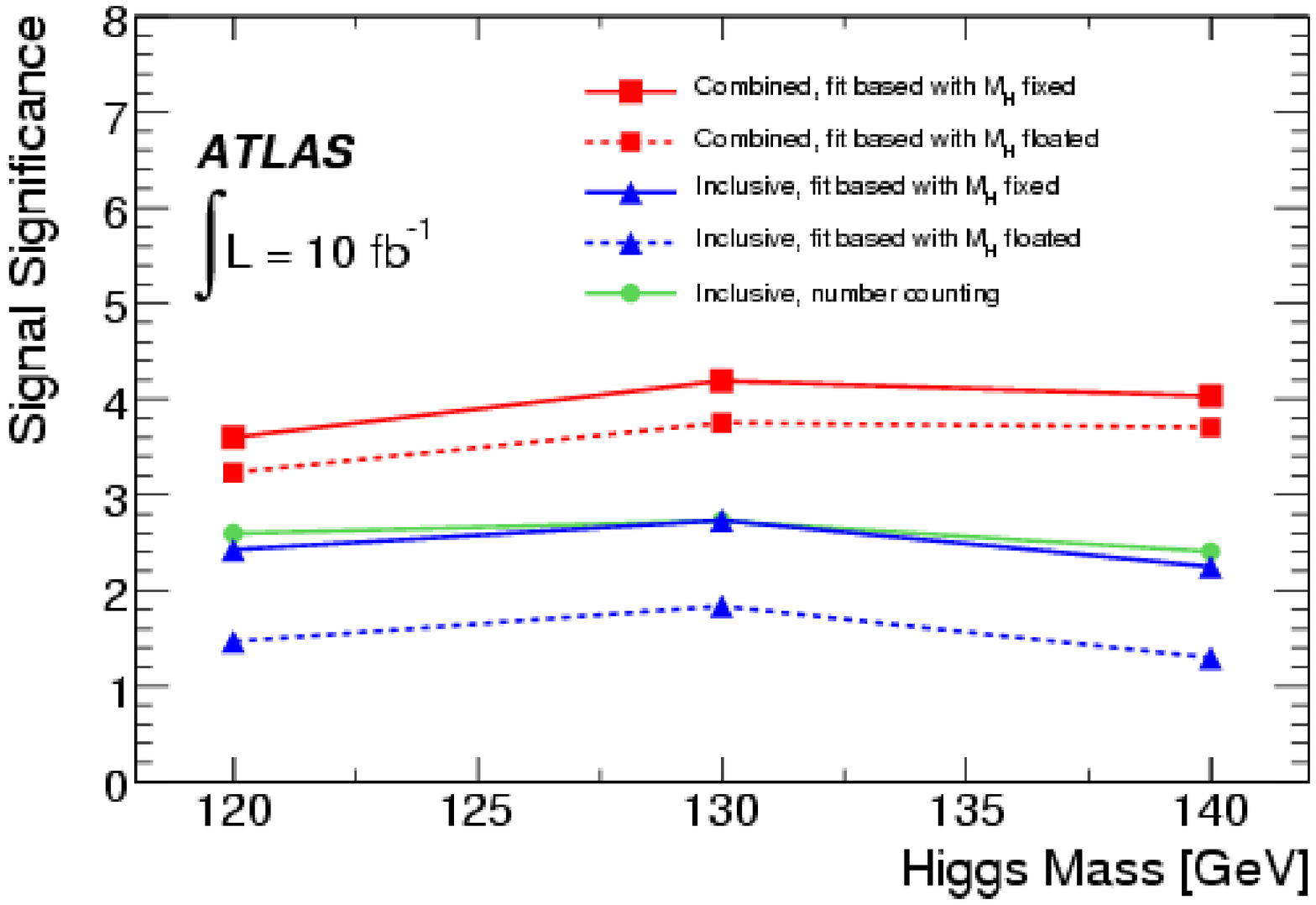} \\
(a) & (b) \\
\end{tabular}
\end{center}
   \caption
       { (a) The invariant mass reconstructed from two high pt photons
	for the inclusive H $\rightarrow \gamma \gamma$ analysis for a
	simple cut based analysis before optimisation done by
	CMS. Different background contributions and signal events with
	4 different masses are shown. (b) The statistical significance
	of the observed events that can be achieved at an integrated
	luminosity of 10~\invfb, which corresponds approximately to one
	year of data-taking at low luminosity of \lowlumi. A
	comparison is made between the inclusive searches and the
	combined results from categorised optimisation.  }
\label{fig:Hgamgam}
\end{figure}

The Higgs decay to a photon pair is one of the most promising channels
for an early discovery of the Higgs boson in the low mass range,
despite its small branching ratio and the large background
contributions expected. The dominant background to this channel with
two real photons arises from the prompt di-photon
production. Additionally there is a large contribution from the
so-called instrumental background, which are gamma + jet or multi-jet
events where one (or two) jet is mis-identified as a photon. The Higgs
boson signal would appear as a resonance peak in the distribution of
the invariant mass of the two photons (M$_{\gamma\gamma}$) above the
background continuum. Figure~\ref{fig:Hgamgam} (a) shows the
reconstructed di-photon mass distributions for background and signal
processes with different mass scenarios in an inclusive
search~\cite{bib:CMS_Hgamgam}. The background contributions in the
signal region are estimated from the side bands of the
M$_{\gamma\gamma}$ distribution.

The narrow width of the signal mass peak is the key to observing such
events over a large background, and an excellent electromagnetic (EM)
calorimeter resolution is crucial. Both ATLAS and CMS detectors have
been carefully designed to detect this one of this bench mark channel,
and to maximise their performance the calibration of the detector
during the early years of data taking become important tasks. For the
CMS experiment, various methods have been developed for the
intercalibration of the uniformly distributed lead tungstate crystals
in the EM calorimeter to achieve the design goal resolution of less
than one percent at high energy. The instrumental background
contributions are largely reduced by applying a tight tracker
isolation and selection based on the shower shape variables upon
photon selection. A significant fraction of the photons convert before
reaching the calorimeter due to the material in the central detector;
a further signal yield can be obtained by identifying and recovering
these converted photons. Complimentary to the lateral shower shape,
ATLAS also uses the longitudinally segmented information of the energy
deposition in the EM calorimeter to identify the fakes and converted
photons.

Both experiments employ multivariate techniques, such as likelihood
and neural network, to increase the sensitivity of the search. The
inputs to obtain the final multivariate discriminant include those
variables developed for photon identification and those based on the
event kinematics: e.g. \pt~of the di-photon system and the photon
decay angle~\footnote{the variable used is cos$\theta^{*}$, 
where $\theta^{*}$ is the photon decay angle in teh Higgs rest frame 
with respect to the Higgs flight direction in the lab frame.}. For 
a further optimisation, events are categorised
depending on the detector region, photon shower shape, jet
multiplicity etc. Recent ATLAS studies show that the signal
significance can be improved by a factor of 1.5 or more by combining
the categories which are separately analysed and optimised
(Fig.~\ref{fig:Hgamgam} (b))~\cite{bib:ATLAS_prelim}.

\subsection{qqH, H\rarr\tautau}
\label{sec:Htautau}

The hadronic decay of tau lepton is distinct from a jet; its final
state contains a limited number of charged particles (1 or 3 tracks
considered for the majority of the analyses) which are relatively
collimated, forming a narrow energy deposition in the calorimeters.
The Higgs decay into two taus becomes a powerful channel when combined
with the Vector Boson Fusion (VBF) production. In addition to the
Higgs decay signal of leptons (electron/muon) and/or hadronic taus,
properties of two very forward out-going quarks from the VBF process
provide further discrimination against other Standard Model
processes. The most dominant background to this channel is the
Z\rarr\tautau~with associated jets, which are mostly produced via QCD
processes involving gluons. The signal VBF events can be distinguished
from those Z+jets background since the two leading quark jets are well
separated in pseudorapidity, and the hadronic activity in the central
region is heavily suppressed due to the colour coherence of the
initial and final state radiation from each of the quarks.  The event
selection is mainly based on the VBF jets: kinematics of individual
jets and angular separation and invariant mass of the di-jet
system. The selection cuts based on the VBF jets are equally effective
against various instrumental/reducible backgrounds such as W+jets,
multi-jet and tt events, and their contributions can be reduced by
orders of magnitude. In addition, events with central hadronic
activities (jets and/or tracks excluding those from leptons or
hadronic tau) are vetoed for a further reduction of background events
by a factor of two while maintaining $\sim$90\% signal
efficiency~\cite{bib:CMS_Htautau}.

\begin{figure}[tb!]
\begin{center}
\begin{tabular}{cc}
\includegraphics[width=.38\textwidth]{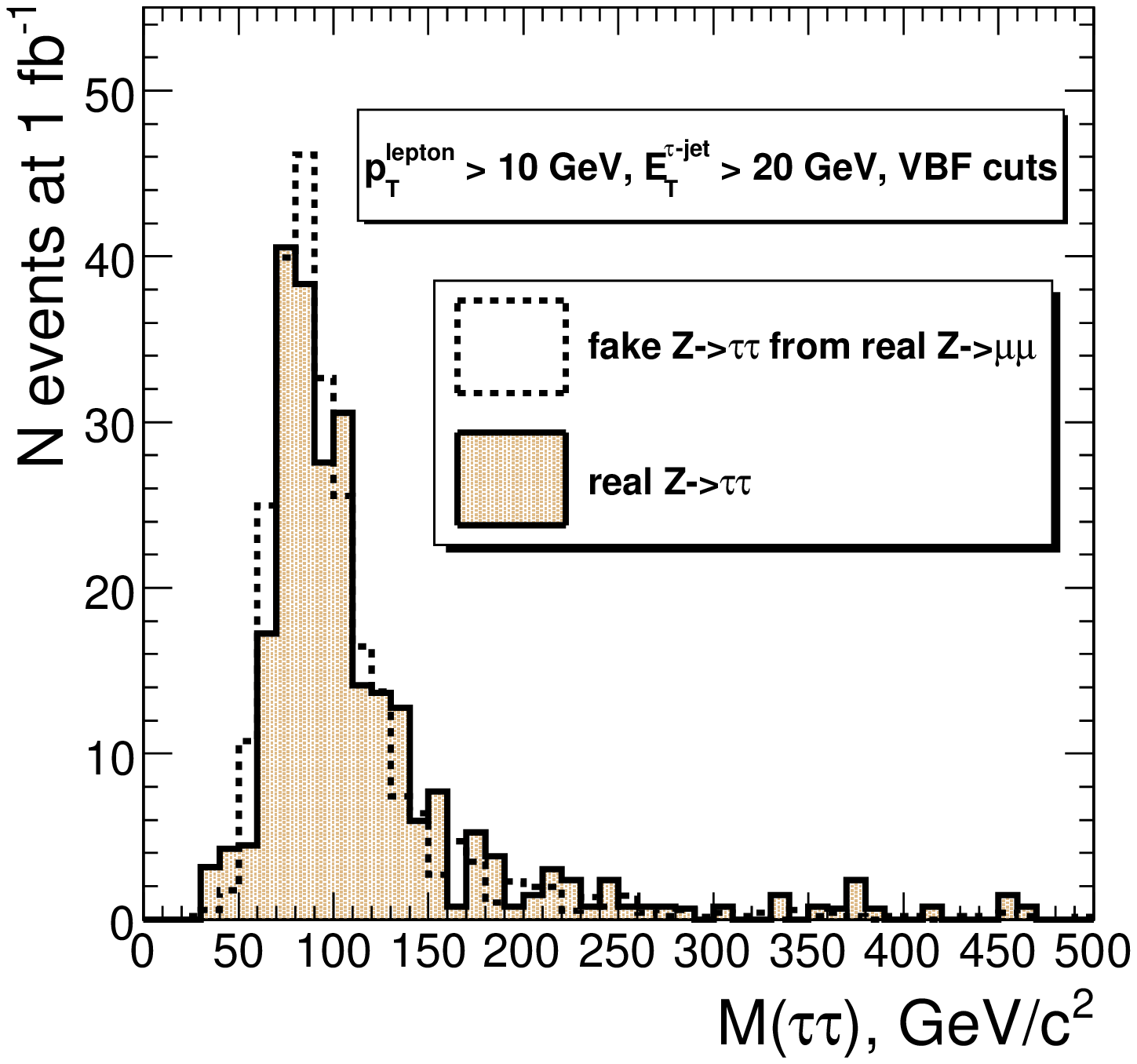} 
& \includegraphics[width=.36\textwidth]{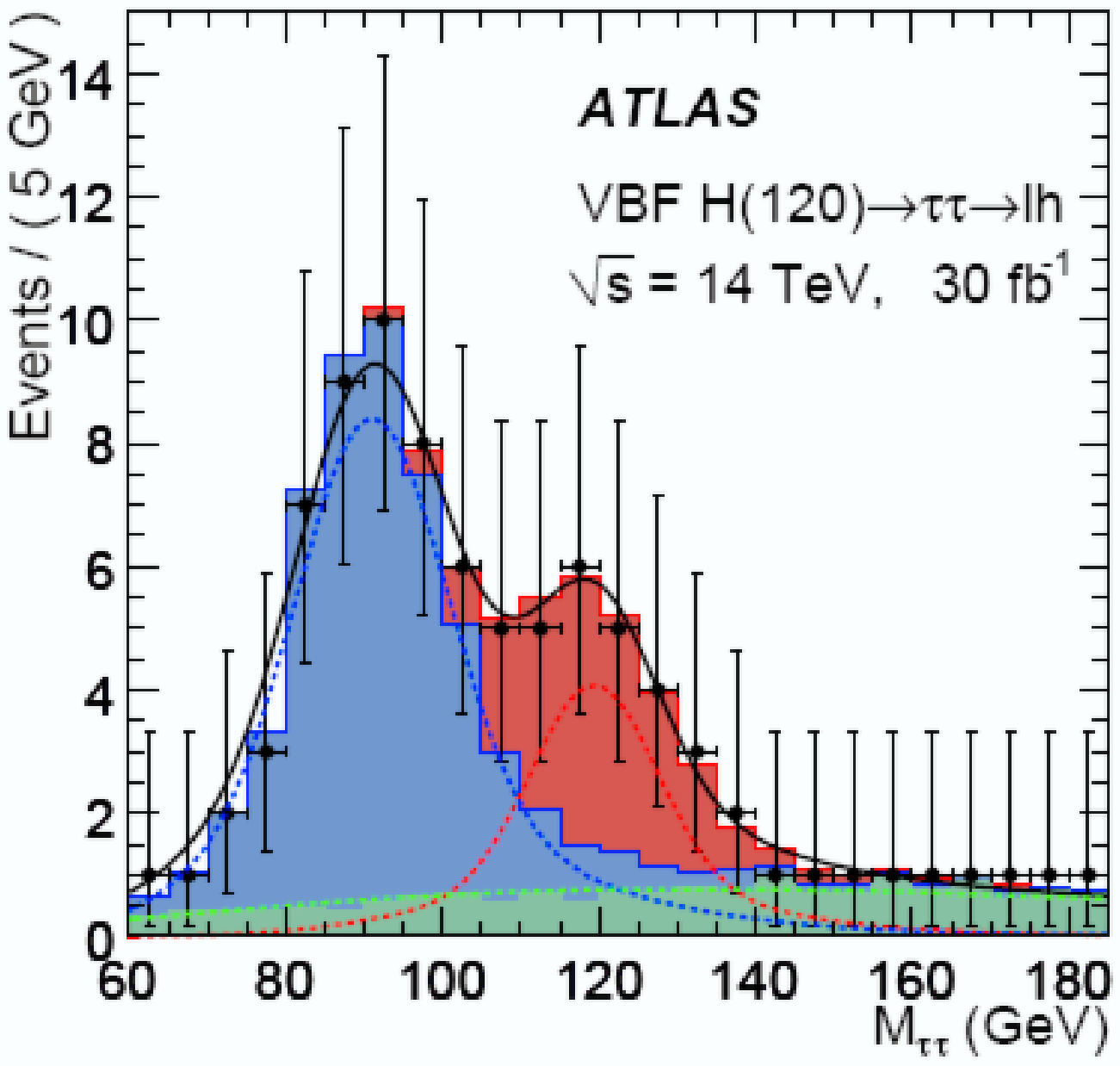} \\
(a) & (b) \\
\end{tabular}
\end{center}
   \caption
       { (a) Demonstration of Z\rarr\tautau~mass shape estimation
	using Z\rarr$\mu\mu$ events done by CMS. A relaxed set of cuts
	are applied on the leading VBF jets, assuming the data in the
	1st year of physics runs at the LHC. The fake (modeled)
	M$_{\tau\tau}$ distribution is in good agreement with the real
	Z\rarr\tautau~events. (b) The full invariant mass of two taus
	reconstructed using collinear approximation for the
	H\rarr\tautau\rarr$lh$ (one tau decaying leptonically and the
	other hadronically) channel (ATLAS). The M$_{\tau\tau}$
	distribution is shown for the mass of 120~\gevcc~and the
	events are normalised to the number expected to be observed
	after all selection cuts at 30~\invfb~of data, which can be
	collected in $\sim$3 years at low luminosity. The mass shape
	is fitted with signal $+$ background hypothesis.}
\label{fig:Htautau}
\end{figure}

The mass of the Higgs boson cannot be directly reconstructed for the
H\rarr\tautau~channel because of the neutrinos involved in the tau decays.
A collinear approximation is used to first reconstruct the full tau
lepton energy by projecting the missing \et~to the direction of the
visible decay products, and then to calculate the mass of the Higgs
boson as the invariant mass of the two tau leptons. Missing
\et~reconstruction involves various parts of the detector thus a
challenging variable to measure at hadron colliders, and some
inefficiency from this method is expected for events with poorly
reconstructed missing \et. However, the collinear approximation also
effectively removes a large fraction of those background processes
with no/small missing \et~such as Z\rarr$ll$ ($l$=e/$\mu$) events.  It
is important to precisely model the shape of the Z\rarr\tautau~mass
distribution, especially if the mass of the Higgs boson is relatively
small that the M$_{\tau\tau}$ resonance peak would appear near the Z mass. It
has been proposed to use the Z\rarr$\mu\mu$ data replacing the
detected muons by the simulated tau decays, such that the rest of the
events (associated jets, minimum bias, underlying events, etc.) are
the same as that in data (Fig.~\ref{fig:Htautau} (a))
~\cite{bib:CMS_Htautau}. The Z\rarr ee background with electrons
mis-identified as hadronic taus is particularly a dangerous background
in the expected signal region. A dedicated variable is developed to
reject electrons which is otherwise identified as hadronic taus, and
their contribution is planned to be estimated from data by using
inverted cuts~\cite{bib:CMS_Htautau}.

An example of the invariant mass distribution of the selected signal
and background events is shown in Figure~\ref{fig:Htautau} (b) after
all the selection cuts are applied~\cite{bib:ATLAS_prelim}. The Higgs
boson signal is clearly visible on the high mass tail of the Z mass
distribution, and the rest of the background is expected to be flat
over the mass range concerned. The events are normalised to the
statistics expected at the integrated luminosity of 30~\invfb,
equivalent to the kind of time-scale that a 5$\sigma$ discovery is
expected to be within the reach for both ATLAS and CMS.

\subsection{ttH, H\rarr bb}
\label{sec:Hbb}

The Higgs decay to bottom quark pair has the largest branching ratio
at low mass. At the same time, it is a very challenging channel to
observe due to its hadronic final state. Jets originated from b-quarks
differ from those from light quarks in characteristics such as charged
particle multiplicity and secondary vertex displacement, hence a
moderate level of background rejection can be achieved by tagging
these b-jets. Nonetheless, the background rate is still high due to
the high cross section of multi-jet events at the LHC. In addition,
the di-jet mass resolution is less defined compared to the mass
reconstructed from leptons because the energy and position resolution
of a jet is generally much worse than the lepton resolution. For this
decay channel, a production via tt fusion has been considered by both
experiments to increase the background rejection power. The weak decay
of the top quarks produces additional b-jets, and the leptonic and/or
hadronic decay of the W bosons also helps to signify the signal
process. The high jet (b-jet) multiplicity and a presence of a lepton
reduce the instrumental background to a negligible level, and the
dominant background becomes the ttbb production which has a cross
section an order of magnitude higher than that of the ttH Higgs
production. The final state with four b-jets introduces a
combinatorial background where a wrong b-jet pair may be associated to
the Higgs decay, which is an additional complication to the analysis
of this channel.

\begin{figure}[tb!]
\begin{center}
\includegraphics[width=.5\textwidth]{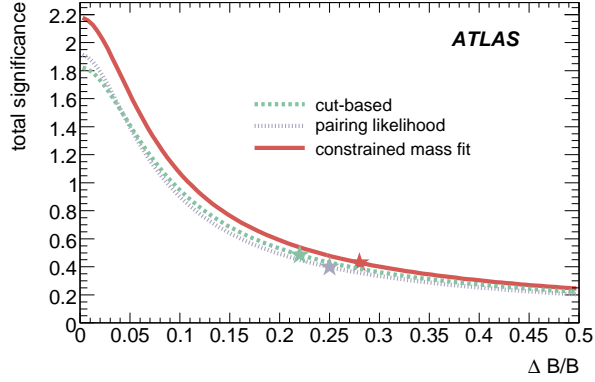} 
\end{center}
   \caption
       { The signal significance in terms of number of sigmas that can
	 be achieved for the search of ttH\rarr ttbb process shown for
	 different level of uncertainty ($\Delta$B) on the number of
	 background events (B). An integrated luminosity of 30~\invfb~is
	 assumed.}
\label{fig:Hbb}
\end{figure}

Due to the large theoretical uncertainty on the background cross
sections, their contributions are planned to be estimated from
data. The shape of the b-jet invariant mass distribution is extracted
from the loosely selected samples, and the absolute normalisation is
obtained from the side-bands in the final selected sample of
events. The control of the experimental uncertainties is a major task;
the largest sources of uncertainties are the jet resolution, jet
energy scale and b-tagging efficiency. Figure~\ref{fig:Hbb} shows the
signal significance that can be achieved as a function of the
uncertainty on the number of background
events~\cite{bib:ATLAS_prelim}. It illustrates that the systematic
uncertainty needs to be understood to a level of few \% in order to
claim greater than 1$\sigma$ significance at integrated luminosity of
30~\invfb~after few years of low luminosity run.

\section{Higher Mass Higgs Searches}

\subsection{H$\rightarrow$WW}
\label{sec:HWW}

The H\rarr WW process becomes most significant in the intermediate
mass region when the Higgs boson is heavy enough to produce the W
bosons on-shell. The discovery reach can be further extended by
performing a dedicated search for the Higgs production via Vector
Boson Fusion where additional background suppression can be
achieved. The leptonic decay of the W bosons is favoured for
triggering reasons as well as the level of the background contributions
expected at the LHC. However, the involvement of neutrinos in the W
decay makes the analysis less straight forward. Due to its heavier
mass, those neutrinos will not be collinear with the leptons as
demonstrated for the H\rarr\tautau~analysis, hence it is almost impossible
to reconstruct a well defined mass peak. The H\rarr WW search is
therefore performed as a counting experiment, where a precise
knowledge of the background contents in data is crucial.

The single most powerful discriminant for the H\rarr WW channel is the
$\phi$ separation of the two leptons from the W decays. Compared to
the dominant WW background from the continuum qq\rarr WW and \ttb, the
W boson (hence the decay leptons) are strongly spin-correlated in the
signal process since they decay from a single Higgs boson with a spin
of zero. This gives rise to a relatively small angular separation
between the final state leptons contrarily to the background WW
production. The events in the high \dphi~region (control region)
can be used to model the background and to estimate their
contributions in the signal region. Other variables used for the
H\rarr WW search include di-lepton mass, WW transverse mass and
missing \et. For the exclusive search for the VBF production, similar
cuts on the kinematic variables based on VBF jets which are introduced
for the H\rarr\tautau~analysis (Section~\ref{sec:Htautau}) are
used. Central Jet Veto is as effective for the inclusive search to
reduce the tt background contributions. In the past studies,
different set of selection cuts are applied and/or relaxed/inverted in
order to separately estimate the different background contributions
(for e.g.~\cite{bib:CMS_HWW}).

\begin{figure}[tb!]
\begin{center}
\begin{tabular}{cc}
\includegraphics[width=.5\textwidth]{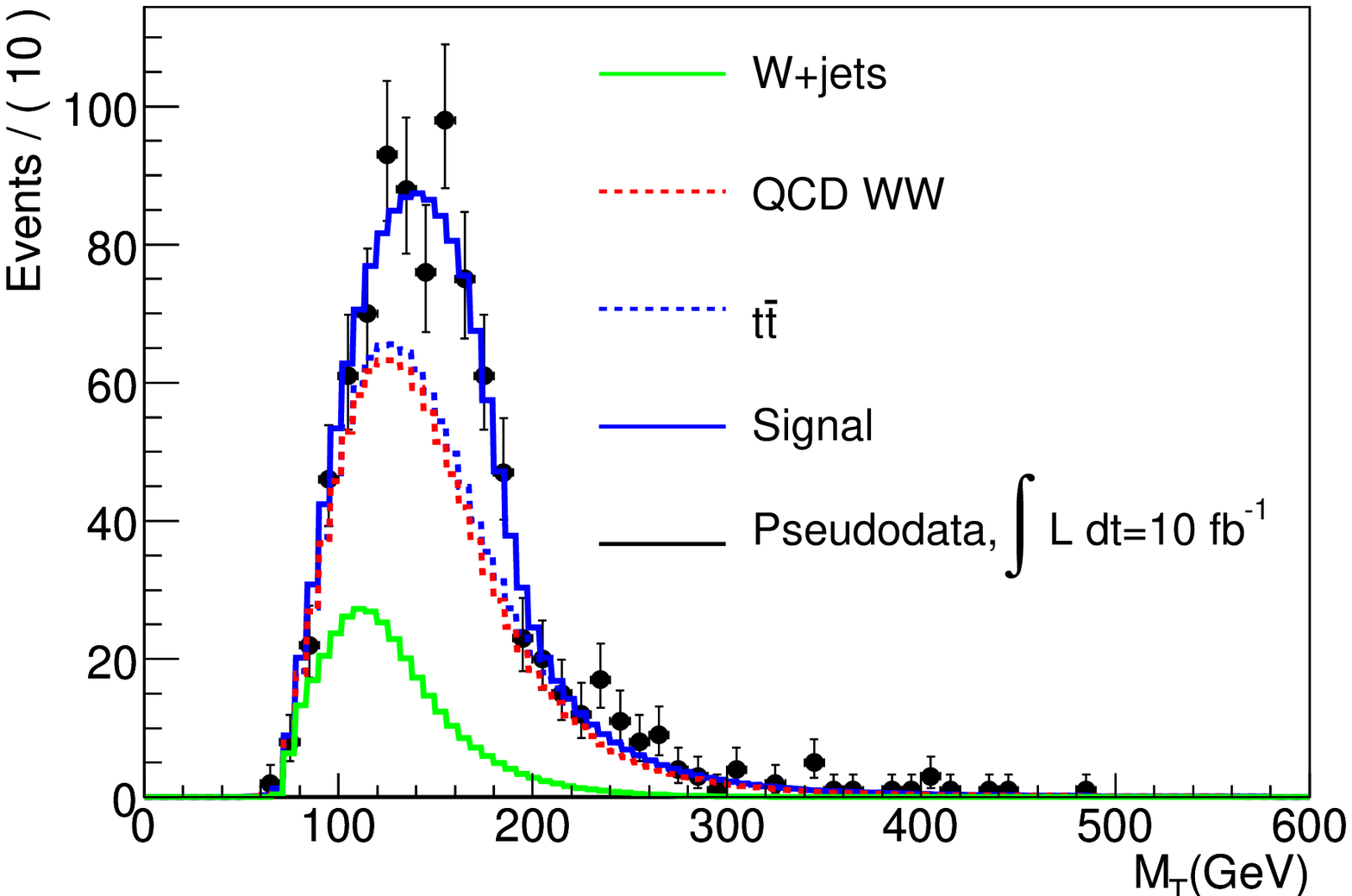} 
& \includegraphics[width=.38\textwidth]{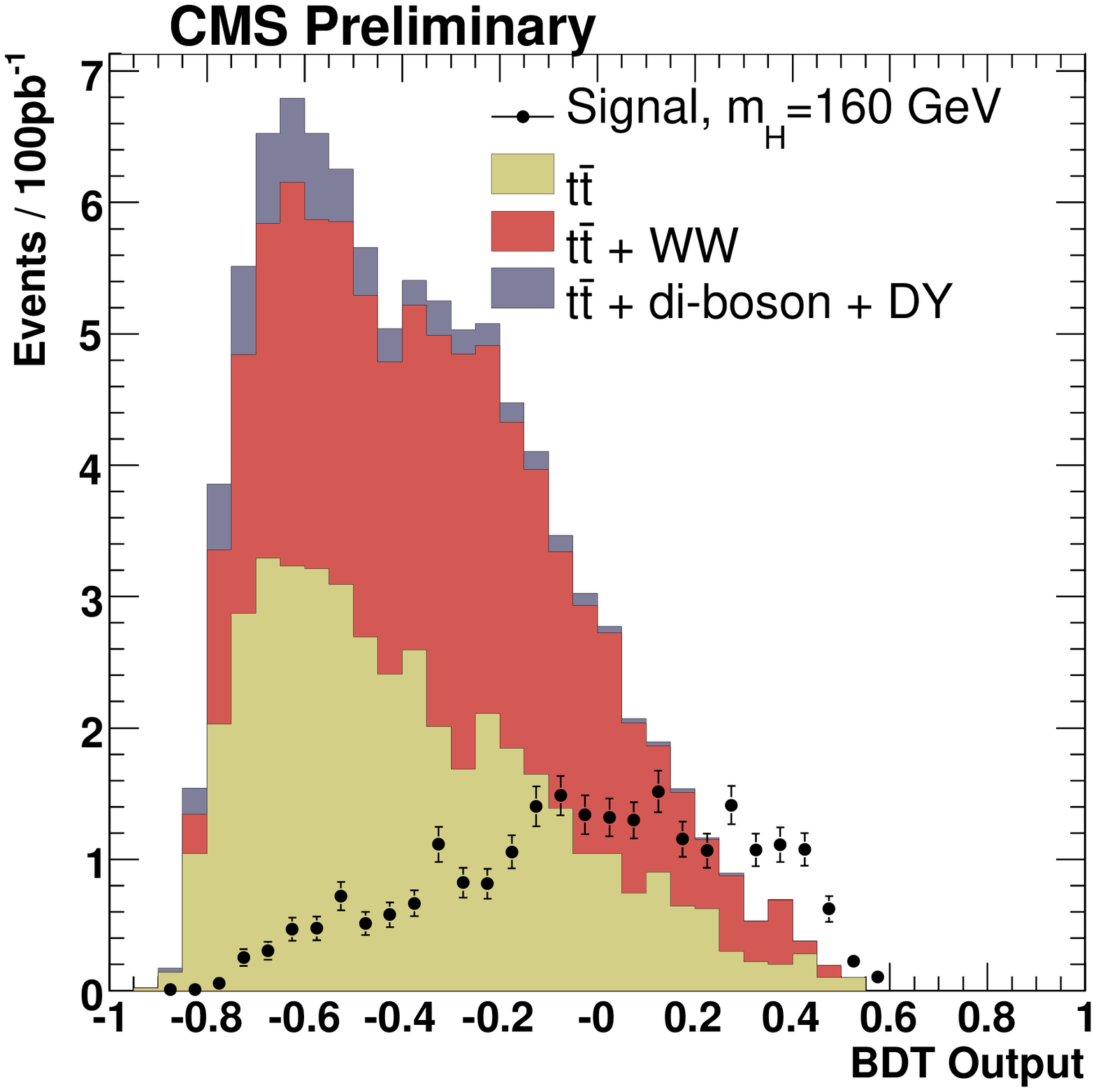} \\
(a) & (b) \\
\end{tabular}
\end{center}
   \caption
       { (a) The transverse mass distribution in the signal region of
	 $\Delta\phi_{ll}$ $<$ 1.575 (ATLAS). The distribution for the
	 signal and background MC events are shown with superimposed
	 pseudodata which are generated for an integrated luminosity
	 of 10~\invfb. (b) The output of the Boosted Decision Tree for
	 the Higgs signal with a mass of 160\gevcc~and various
	 background processes. The events are normalised to
	 100~\invpb~of integrated luminosity, which corresponds to the
	 data that can be collected in the first months of the physics
	 data taking at the LHC.}
\label{fig:HWW}
\end{figure}

The signal and background contents in the signal region of small
di-lepton separation are shown in the transverse mass distribution
after selection cuts are applied in Figure~\ref{fig:HWW} (a)
~\cite{bib:ATLAS_prelim}. Already at 10~\invfb, a significant excess
from the H\rarr WW signal is expected to be observed at the level of
tens of \% of the number of estimated background events from other
Standard Model processes. A sizable systematic uncertainty of
$\sim$15\% arises from the background estimation, and it may increase
up to $\sim$50\% with very early data due to the limited
statistics. The multivariate techniques are used to increase the
sensitivity of the search at the relatively early stage of the physics
runs. Figure~\ref{fig:HWW} (b) shows an example of the output from a
multivariate analysis using Boosted Decision Trees (BDT). Several
kinematic variables including \dphi~are used as the inputs to the
BDT~\cite{bib:CMS_HWWprelim}. A moderate separation between the signal
and the background events is achieved, enhancing the signal
significance in the positive output value. A correct usage of
multivariate tools requires sufficient understanding of all the input
variables, and a good description of the background distribution in
variables involving jets and missing \et~remains an important task.

\subsection{H$\rightarrow$ZZ}
\label{sec:HZZ}

The H\rarr ZZ decay is often referred to as the "Golden Channel" at the
LHC because of its very clear signature of multiple high \pt~leptons
(electron/muon) without any hadronic objects or missing \et~involved in
the final state. This decay mode has a significant advantage over
other channels in terms of branching ratio at high mass, and the
sensitivity of the search is further enhanced by the relatively small
background contributions. Both ATLAS and CMS detectors are designed to
deliver an excellent electron and muon identification and
reconstruction performance for this benchmark channel. A particular
effort has been put into the reconstruction of low \pt~leptons down to
few \gevc~in order to access the low mass region.

The event selection is mainly based on the lepton quality and
kinematics. The vertex information is used to reconstruct the
invariant mass of the leptons. The dominant sources of background to
the 4-lepton signal are the Zbb, tt and ZZ production. Some of the
leptons in the Zbb and tt processes are produced with b-jets and can
be suppressed by requiring a tight isolation on these leptons. A
further reduction of the background is achieved by associating the
leptons to Z. The oppositely charged leptons are paired and their
invariant mass is required to fall within the Z mass window. The best
pairs are used for 4e and 4mu channels where more than one combination
is possible. The contributions from the remaining ZZ background is
planned to be obtained from data: either by fitting the known ZZ shape
to the sidebands of the 4-lepton invariant mass distribution or by normalising it to the
cross section predicted by the single Z production cross section. The
former method can achieve better precision when there is sufficient
statistics of data, however, in the early period of data-taking when
the discovery of the Higgs boson through this channel is already
possible it suffers from a large statistical uncertainty. For the
second method, the Z production cross section can be measured to a
precision at an integrated luminosity of the order of
$\sim$10~\invfb. Taking the ratio to the ZZ cross sections cancels out
some of the theoretical and experimental uncertainties. The overall
uncertainty on the background events for both methods is expected to
be at the level of $\sim$ few \%.

\begin{figure}[tb!]
\begin{center}
\begin{tabular}{cc}
\includegraphics[width=.5\textwidth]{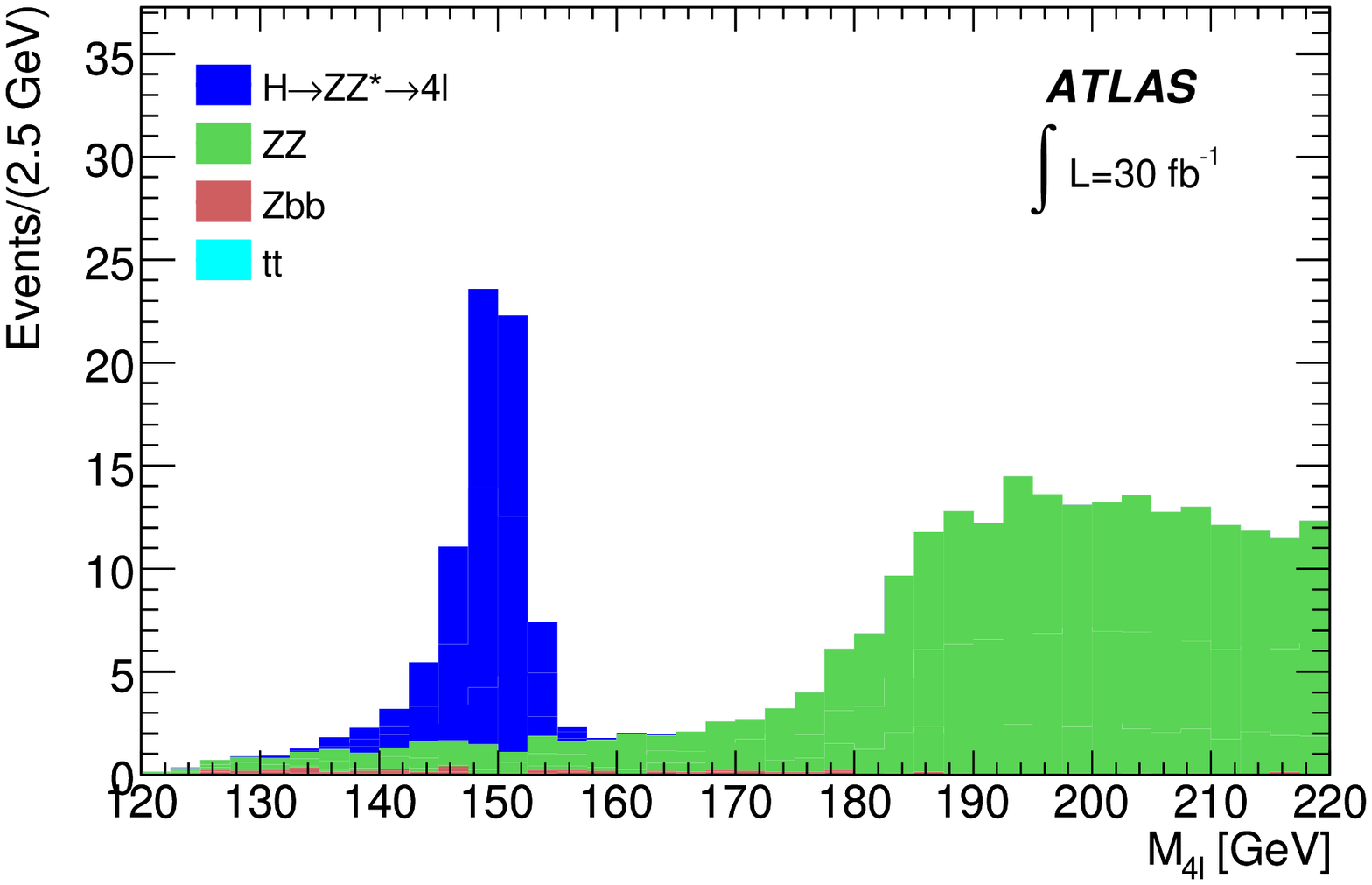} 
& \includegraphics[width=.5\textwidth]{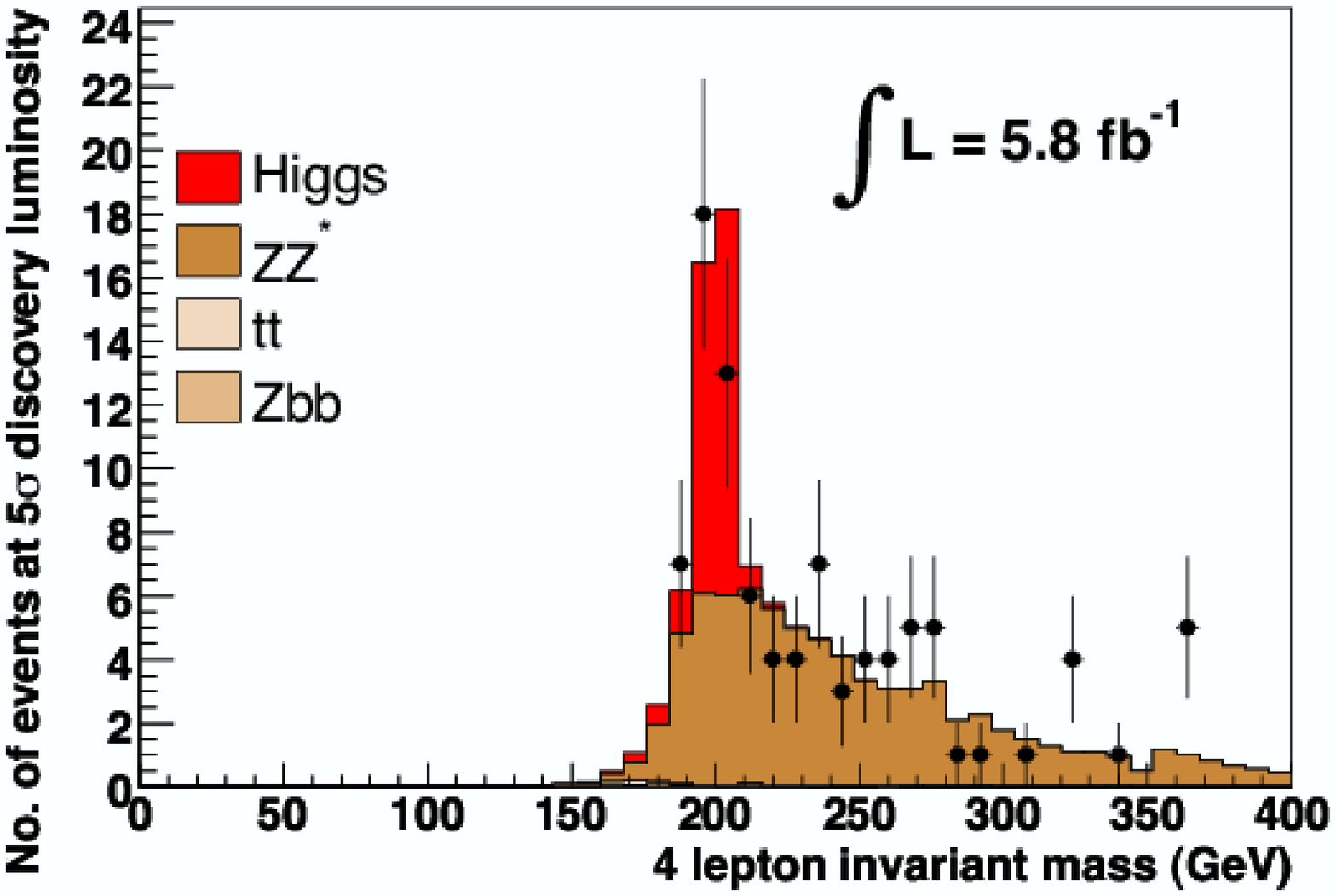} \\
(a) & (b) \\
\end{tabular}
\end{center}
   \caption
       { (a) The invariant mass of the four leptons for a Higgs mass
	 of 150~\gevcc. For the signal process, one of the Z boson is
	 produced off-shell. The events are normalised to an
	 integrated luminosity of 10~\invfb, which corresponds to
	 $\sim$one year of low luminosity run. (b) The M$_{\rm 4 l}$
	 distribution for the H\rarr ZZ\rarr ee$\mu\mu$ channel with the
	 statistics need for 5$\sigma$ discovery (CMS). The MC signal and
	 background processes are shown with pseudo data randomly
	 generated according to the predicted distribution.}
\label{fig:HZZ}
\end{figure}

The reconstructed invariant mass of the four leptons is shown in
Figure~\ref{fig:HZZ} (a) after the selection for a generated Higgs
mass of 150\gevcc~\cite{bib:ATLAS_prelim}. The events are normalised to the
expected number of events at the integrated luminosity of
10~\invfb. In this relatively low mass range, the ZZ background level
is very low since the production of real (on-shell) Z boson pair is
suppressed below the threshold of $\sim$twice the mass of
Z. Figure~\ref{fig:HZZ} (b) shows the M$_{\rm 4 l}$ distribution
normalised to the statistics needed to achieve 5$\sigma$ discovery
for a high mass Higgs boson of 200~\gevcc~\cite{bib:CMS_PTDR}. The particular
decay mode of ZZ\rarr ee$\mu\mu$ has advantages over the 4e or 4$\mu$
channels by the fact that the branching ratio is doubled. It also
benefits from the absence of the combinatorial background since the
leptons can be only paired within the same flavour. For both plots of
the 4-lepton invariant mass distribution, the higher order corrections are
applied to the background cross section.

\section{Discovery Potential}
\label{sec:reach}

\begin{figure}[tb!]
\begin{center}
\begin{tabular}{cc}
\includegraphics[width=.4\textwidth]{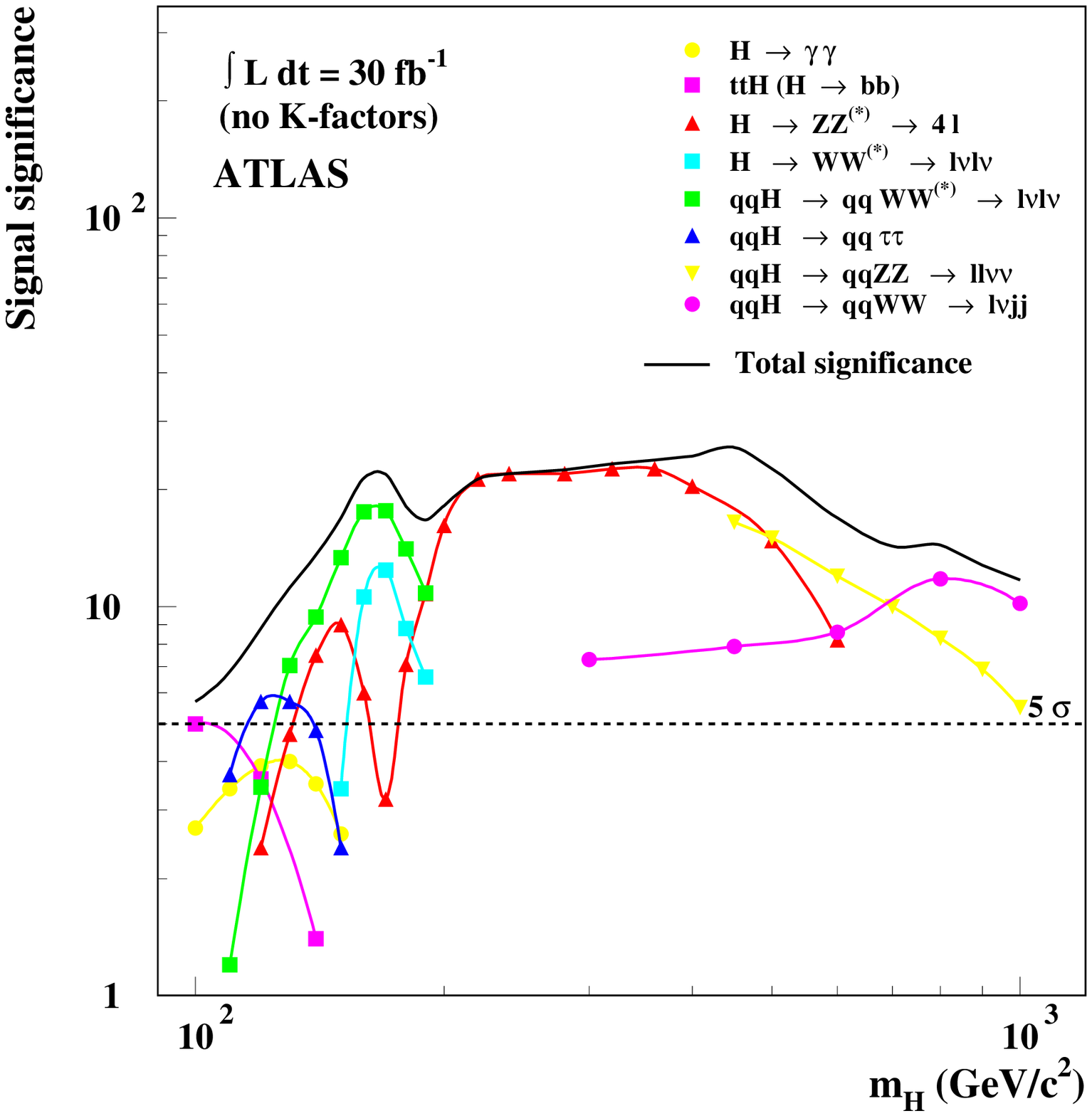} 
& \includegraphics[width=.4\textwidth]{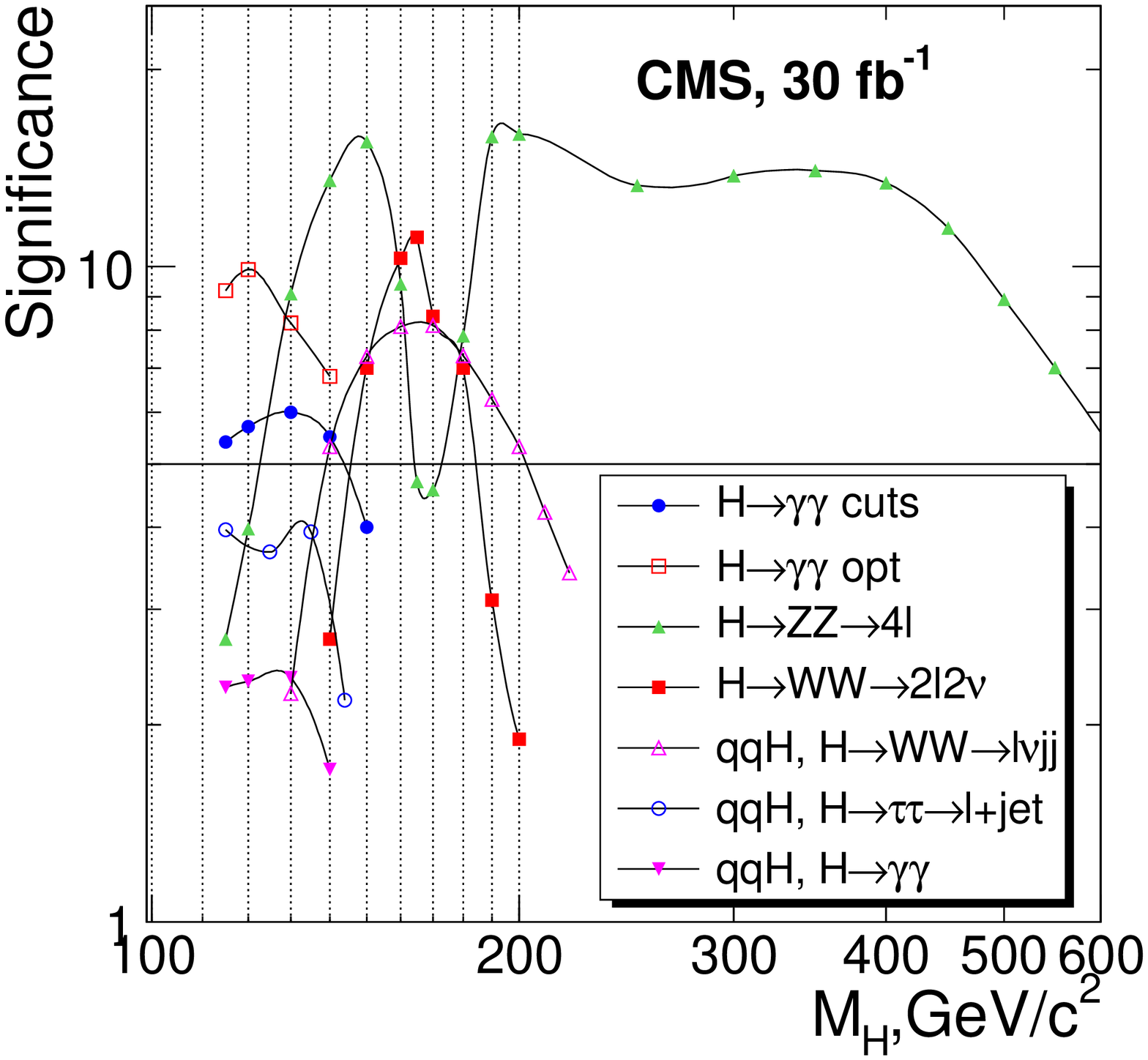} \\
(a) & (b) \\
\end{tabular}
\end{center}
   \caption
       { The significance of the signal events expected to be observed
	 by the (a) ATLAS and (b) CMS experiments for a range of Higgs
	 boson mass with an integrated luminosity of 30~\invfb, which
	 corresponds to $\sim$3 years of data taken with the low
	 luminosity runs (L = \lowlumi).}
\label{fig:reach}
\end{figure}

The Higgs boson discovery reach of the ATLAS and CMS experiments at
the integrated luminosity of 30~\invfb~are summarised in
Figures~\ref{fig:reach} (a) and (b) for the entire range of Higgs
boson mass discussed~\cite{bib:ATLAS_reach}~\cite{bib:CMS_PTDR}. For
ATLAS, the low to intermediate mass range is well covered by the
H\rarr\tautau~and H\rarr WW decay channels combined with the Vector
Boson Fusion production, reflecting the strength of the ATLAS detector
in the jet and missing \et~reconstruction. On the other hand, the
H\rarr\gamgam~decay channel is most promising for CMS where it
benefits from the high performance EM calorimeter, and can achieve
5$\sigma$ significance at Higgs masses down to $\sim$110~\gevcc. The
two experiments provide similar performance in the high mass region,
and ATLAS further extends the reach by including the VBF production
process. With 30~\invfb~of data, the discovery of the Higgs boson is
possible for all the mass range up to the TeV scale by each of the
experiments alone. There has not been a recent combination of results,
however, with ATLAS and CMS together the early discovery is within the
reach.

\section{Summary}

The prospects for the Standard Model Higgs boson search at the LHC
have been presented. A wide range of Higgs production and decay
channels are studied by the two general purpose detector experiments,
ATLAS and CMS, and the results show that the discovery would be
possible for all the mass range already in the first year or two of
physics data. There are continuing efforts to improve the analyses and
to be ready for the first physics data at the LHC expected in the
coming year.


\begin{thebibliography}{9}   

\bibitem{bib:LEP}
The LEP Working Group for Higgs Boson Searches (ALEPH, DELPHI, L3 and
OPAL Collaborations), ``Search for the Standard Model Higgs Boson at LEP'', 
Phys. Lett. B 565 (2003) 61, hep-ex/0306033, CERN-EP/2003-011

\bibitem{bib:CMS_Hgamgam} 
M. Pieri, S. Bhattacharya, I. Fisk, J. Letts, V. Litvin, J.G. Branson,
``Inclusive Search for the Higgs Boson in the H\rarr\gamgam~Channel'', 
CMS NOTE 2006/112.

\bibitem{bib:ATLAS_prelim} 
ATLAS Collaboration,
``Expected Performance of the ATLAS Experiment, Detector, Trigger and Physics'',
CERN-OPEN-2008-020, Geneva, 2008, to appear.

\bibitem{bib:CMS_Htautau}
CMS Collaboration, 
``Towards the search for the Standard Model Higgs boson produced in 
Vector Boson Fusion and decaying into a tau pair in CMS with 1~\invfb: 
tau identification studies'', 
CMS PAS HIG-08-001.

\bibitem{bib:CMS_HWW}
G. Davatz, M. Dittmar, A.S. Giolo-Nicollerat, 
``Standard Model Higgs Discovery Potential of CMS in the H\rarr WW\rarr$l\nu l\nu$ Channel'',
CMS NOTE 2006/047.

\bibitem{bib:CMS_HWWprelim}
CMS Collaboration,
``Search for the Higgs boson in the WW(*) decay channel with the CMS experiment'',
CMS PAS HIG-07-001.

\bibitem{bib:CMS_PTDR} 
CMS Collaboration,
``CMS Physics Technical Design Report Volume II: Physics Performance'',
CERN/LHCC/2006-021, CMS TDR 8.2.

\bibitem{bib:ATLAS_reach} 
K. Cranmer, B. Mellando, W. Quayle, S. L. Wu, 
``Statistical Methods to Assess the Combined Sensitivity of the ATLAS Detector
to the Higgs Boson in the Standard Model'',
ATL-PHYS-2004-034.

\end{thebibliography}
\end{document}